\documentclass[runningheads]{llncs}

\usepackage{graphicx}
\begin{document}
\title{Detecting brute-force attacks \\ on cryptocurrency wallets~\thanks{Supported by the Russian Foundation for Basic Research (18-37-20033).}}

\author
{
	E.O. Kiktenko~\inst{1}\and
	M.A. Kudinov~\inst{1,2}\and  
	A.K. Fedorov~\inst{1} 
}
\institute{
	$^1$Russian Quantum Center, Skolkovo, Moscow 143025, Russia\\
	$^2$Bauman Moscow State Technical University, Moscow 105005, Russia\\
}

\newcommand{\alert}[1]{{\color{blue} #1}}

\maketitle             
\begin{abstract}
Blockchain is a distributed ledger, which is protected against malicious modifications by means of cryptographic tools, e.g. digital signatures and hash functions. 
One of the most prominent applications of blockchains is cryptocurrencies, such as Bitcoin. 
In this work, we consider a particular attack on wallets for collecting assets in a cryptocurrency network based on brute-force search attacks.
Using Bitcoin as an example, we demonstrate that if the attack is implemented successfully, a legitimate user is able to prove that fact of this attack with a high probability.
We also consider two options for modification of existing cryptocurrency protocols for dealing with this type of attacks.
First, we discuss a modification that requires introducing changes in the Bitcoin protocol and allows diminishing the motivation to attack wallets.
Second, an alternative option is the construction of special smart-contracts, which reward the users for providing evidence of the brute-force attack.
The execution of this smart-contract can work as an automatic alarm that the employed cryptographic mechanisms, and (particularly) hash functions, have an evident vulnerability.
\end{abstract}

\section{Introduction}

Recently, peer-to-peer payment systems based on the blockchain technology, so-called cryptocurrencies, attracted a significant deal of interest~\cite{Swan2015}. 
The crucial feature of cryptocurrencies is their possibility to operate without a central authority that governs the system.
This becomes possible thanks to the use of specific cryptographic tools inside blockchains, such as digital signatures and hash functions~\cite{Witte2016}.
This type of cryptographic primitives is based on so-called one-way functions.
These are believed to be straightforward to run on a conventional computer but difficult (or practically impossible) to calculate in reverse~\cite{Bernstein2017}. 
For example, multiplying two large prime numbers is easy, but finding the prime factors of a given product is hard. 
Existing algorithms based on such a paradigm are known already more than 40 years, but their security remains unproven. 

A particular element of blockchain is a cryptographic hash function, which is a compressive transformation that takes a string of arbitrary length and reduces it to one of predefined finite length.
It is assumed that the task of inverting hash function is extremely difficult for modern computers. 
Specifically, the mechanism of hash functions enables achieving consensus in the concept know as proof-of-work, which is used in Bitcoin and other cryptocurrencies. 
Another use-case for hash functions in the Bitcoin network for additional security of wallets that are used by network members for collecting assets. 

However, the security status of cryptographic tools may change with time.
In particular, quantum computing, which is a tool for information processing with the use of quantum phenomena such as superposition and entanglement, allows solving particular classes of tasks more efficient than with the use of existing classical algorithms. 
These tasks include integer factorization and discrete logarithm problems or searching for an unsorted database.
From the viewpoint of blockchain security~\cite{Fedorov2018}, the latter can be used for achieving a quadratic speedup in calculating the inverse hash function~\cite{Grover1996}. 
We note that the security of Bitcoin from the viewpoint of attacks from quantum computers has been considered in Ref.~\cite{Tomamichel2018}. 
Large-scale quantum computers that enable to realize such an algorithm, however, are not yet available. 
Meantime, significant efforts of the community are concentrated on the distributed brute-force-type analysis of hash function, for example, in the Large Bitcoin Collider (LBC) project~\cite{LBC}.
It is then important to consider possible practical attacks on hash function in a middle-term horizon. 

In this work, we consider a particular attack on cryptocurrency wallets based brute-force search for digital signature secret keys that match addresses of existing wallets.
We show that the success of the attack can be proven with high probability and provide a lower bound for this probability.
We also suggest a modification of cryptocurrency networks that makes such an attack unprofitable and also consider a mechanism that motivates users to announce the fact of hacking their wallets by brute-force-type attack.
These methods can be employed for creating an additional security level for cryptocurrency network with the possibility to reveal and prove the fact of specific types of malicious transactions.
Our ideas are applicable for a general cryptocurrency model, however, we restrict our consideration to the Bitcoin network. 
We expect that the generalization for other cryptocurrencies can be rather straightforward.

The paper is organized as follows.
In Sec.~\ref{sec:attack}, we describe a brute-force-type attack on cryptocurrency wallets. 
In Sec.~\ref{sec:proof}, we demonstrate that if the attack has been processed successfully then the fact of the brute-force-type attack can be revealed with high probability. 
We also obtain a lower bound for the probability of proving the attack (obtaining the `evidence of the attack'). 
In Sec.~\ref{sec:modifications}, we suggest a modification of consensus rules that makes such an attack unprofitable. 
The proposed mechanism is based on `freezing' potentially stolen assets.  
In Sec.~\ref{sec:reward}, we demonstrate that it is possible to create a specific type of altruistic transactions and corresponding smart-contract allowing honest users to collect assets 
if he/she proofs that the assets from the wallet have been stolen in the result of the brute-force-type attack.
This creates a mechanism that motivates users to announce the fact of hacking wallets by brute-force-type attacks in the blockchain network.
We then conclude in Sec.~\ref{sec:concl}. 

\section{Brute-force attack on Bitcoin wallets} \label{sec:attack}

We start our consideration of a brute-force attack with brief revising the construction of Bitcoin wallets.
Each Bitcoin wallet could be considered a set of \emph{addresses}, which are unique identifiers that are assigned to the possession of certain funds.
The ownership of an address by a person corresponds to his knowledge of the secret (also known as private) key that is used for the construction of a corresponding address.
We note that one person can have unlimited addresses. 
Moreover, for the security reasons it is recommended to generate new address after each withdrawing funds from an existing address.

\subsection{Address generation}\label{}

\begin{figure}
	\centering
	\includegraphics[width=0.5\linewidth]{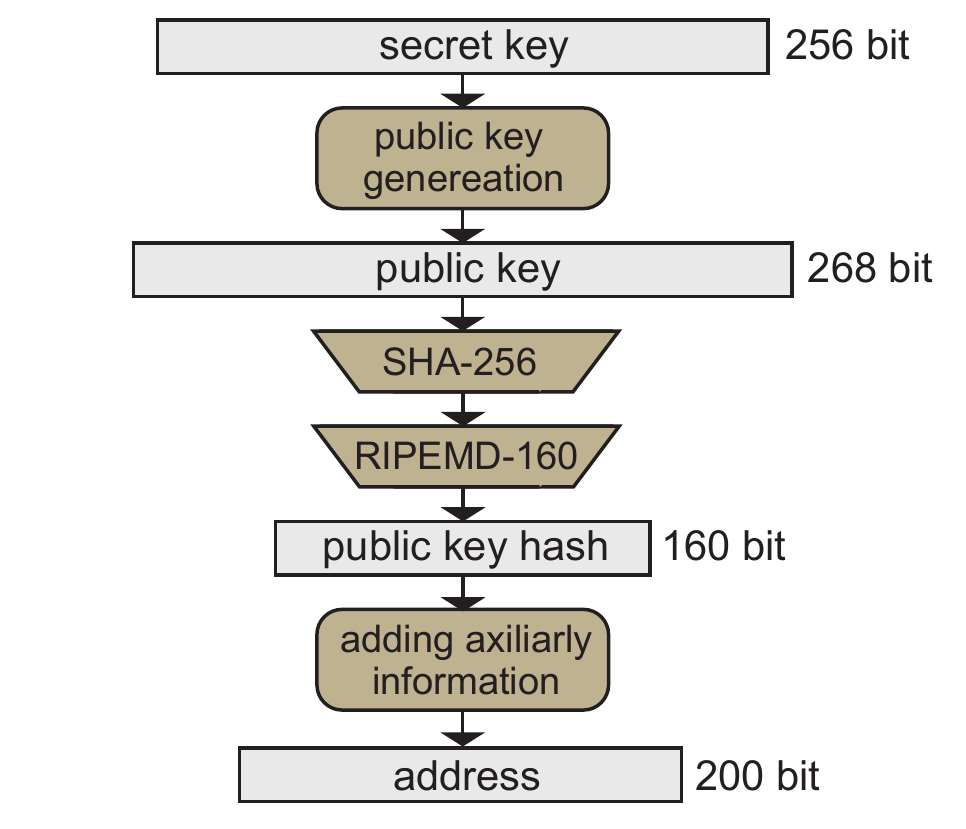}
	\caption{Basic algorithm of the Bitcoin address generation.}
	\label{fig:addrgen}
\end{figure}

The construction of Bitcoin addresses is presented in Fig.~\ref{fig:addrgen}. 
It is based on employing elliptic curve digital signature algorithm (ECDSA)~\cite{ECDSA} defined with secp256k1 standard~\cite{ECDSAstandard}.
The address construction begins with a random generation of a secret key, which is a random string of 256 bits length.
Despite a small restriction on possible secret keys values, almost every 256-bit string corresponds to a valid ECDSA secret key.
Thus, the total number of possible secret keys is given by $N_{\rm sec}\simeq 2^{256}\approx1.16\times10^{77}$.
Then the public key is calculated using the given secret key that is a computationally simple task for modern computing devices. 
We note that the reverse operation of obtaining a secret key from the corresponding known public one is assumed to be computationally infeasible for existing computers.

\newcommand{\SHA}{\textsf{SHA-256}}
\newcommand{\RIPEMD}{\textsf{RIPEMD-160}}
\newcommand{\BASE}{\textsf{Base58Check}}
\newcommand{\PtoPKH}{\textsf{P2PKH}}

The compressed version of the public key, that is a 268-bit string, then goes through irreversible operations of hashing first with \SHA~and then with \RIPEMD~cryptographic hash functions.
As a result, one obtains a 160-bit string that is an essence of Bitcoin address. 
In practice, this hash is concatenated with 4 bytes of a checksum (obtained with doubled \SHA) and an additional version byte, and then the 25-byte (200 bit) result is represented with \BASE~encoding~\cite{Encoding} to obtain a final Bitcoin address.
We note that the transition between \RIPEMD~hash and 200-bit address can be easily reversed.

Due to the intrinsic proprieties of cryptographic hash functions, which make them operates as pseudo-random functions, the effective size of the address space is given by $N_{\rm addr}\simeq 2^{160}\approx1.46\times10^{48}$ that is far less than space size of the possible keyspace $N_{\rm sec}$.
This feature plays a central role in our consideration.
Also, we remind that the security of employed cryptographic hash functions must make it computationally infeasible to find \emph{any} valid public key which gives a particular 160-bit hash.

In what follows we consider only Pay-to-PublicKey Hash (\PtoPKH) transactions.
The idea behind this kind of transactions is that if a user would like to transfer some funds to a particular address, he/she publishes a transaction of a special form with an output containing the given address.
In order to redeem the funds, the owner of the address has to create another transaction with input containing a public key, which hashes to the given address, and a signature, which is created by using the secret key and corresponding public key.
This signature serves as a proof to the system that the author of the transaction is indeed the owner of the address.

\subsection{Attack description}

Here we consider a particular type of a brute-force-type attack.
The considered attack is based on the exhaustive search strategy for the signature forgery attack that is performed by an adversary intended to steal bitcoins from honest users.

\newcommand{\AddrList}{\textsf{AddrList}}
\newcommand{\SKList}{\textsf{SKList}}
\newcommand{\UTXO}{\textsf{UTXO}}

\begin{enumerate}
	\item The adversary makes a list \AddrList~of all the Bitcoin addresses which posses some funds, using the current set of unspent transaction outputs (\UTXO) corresponded to the current state of the Bitcoin blockchain (we remind that we consider \PtoPKH~transactions only).
	\item The adversary chooses a subset \SKList~in the space of all secret keys. This subset is then used for checking over.
	\item\label{itm:mainstep} The adversary takes a secret key from \SKList, generates a corresponding address, and looks whether the obtained address is inside the set \AddrList.
	If this is the case, then the adversary publishes a transaction, which transfers available funds from this address to some pre-generated address owned by the adversary.
	This operation is repeated for all the secret keys in \SKList.
\end{enumerate}

The average rate of finding valid secret keys by the considered attack can be calculated as follows:
\begin{equation}\label{eq:attackrate}
	R = \frac{|\AddrList|}{N_{\rm addr}}R_0,
\end{equation}
where $|\AddrList|$ stands for the size of \AddrList~and $R_0$ is the rate of repeating the step~\ref{itm:mainstep} of the attack algorithm.

We note that here we described only a general idea of the attack.
Some additional steps including updates of \AddrList~can be also considered.
Also, we would like to point out that this attack is implemented against all the users possessing bitcoins rather than a particular user or address.
From Eq.~(\ref{eq:attackrate}) one can see that this fact increases the probability of the success in $|\AddrList|$ times, that is at least a total number of people having Bitcoins at the current moment.

\section{Proving the fact of a successful attack}\label{sec:proof}

\begin{figure}
	\centering
	\includegraphics[width=0.7\linewidth]{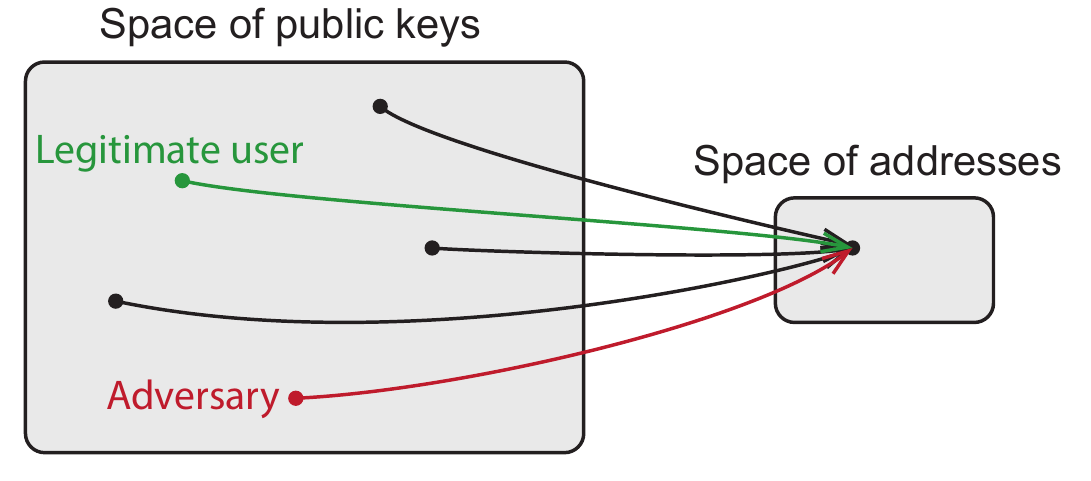}
	\caption{Basic idea behind the providing an evidence of successful brute-force attack: With a high probability the public key of legitimate user will be different from the presented public key of an adversary.}
	\label{fig:collision}
\end{figure}

We consider a scenario, where an adversary succeeded in implementing the brute-force attack considered above and transferred funds from an address of a legitimate user to some another address. 
We claim that this scenario drastically differs from a scenario, where an adversary succeeded in stealing funds by unauthorized access to the secret key of a legitimate user.
The reason is a huge difference between the sizes of secret key space $N_{\rm sec}$ and address space $N_{\rm addr}$.
It turns out that the same address can be generated to approximately $N_{\rm sec}/N_{\rm addr}\sim 10^{29}$ different secret keys. 
Due to the fact that legitimate users (commonly) choose their secret keys at random, in the case of successful attack with very high probability, the secret key chosen by adversary will be different from the one belonging to the legitimate user.
This fact, in turn, leads to the difference between corresponding public keys.
Finally, a successful brute-force attack yields a revealing of collision for \RIPEMD(\SHA($\cdot$)) function (see illustration in Fig.~\ref{fig:collision}).

The important point is that the stealing transaction, which transfers funds from the legitimate user address, essentially includes a public key that hash gives this address.
That is why the legitimate user, possessing another public key, can in principle prove the fact of the successful brute-force by publishing his alternative public key together with the already published public key from the stealing transaction.
The observation of such the collision any user can make sure that there was someone in the network who spent a huge amount of computational resources in order to find a valid secret key for the existing address.
We call the alternative public key, which gives the same address as some another public key, the \emph{evidence of the brute-force attack} (or the evidence for short).

Let us derive a rigorous estimation on the probability that in case of the successful brute-force attack the evidence could be constructed.
In our consideration, we use the following assumptions.
The first is that a legitimate user's secret key is generated with uniformly random distribution over the whole secret keyspace.
The second is that the action of \RIPEMD(\SHA($\cdot$)) hashing satisfies the random oracle assumption~\cite{RandomOracle}: Calculating a hash for every new argument can be considered to be equivalent to a generation of a uniformly random variable in the space of 160-bit strings.

We then consider an address and a corresponding subspace of secret keys, which gives this address.
Under the assumptions described above, the size of this space is given by a random variable as follows:
\begin{equation}
	N \sim {\rm Bin}(N_{\rm sec},1/N_{\rm addr}),
\end{equation}
where ${\rm Bin}(n,p)$ stands for the binomial distribution with number of trials $n$ and success probability $p$.
The probability $\varepsilon$ to obtain the same secret key as the original secret key during the brute-force attack is as follows: 
\begin{equation}\label{eq:eps}
	\varepsilon = \sum_{n=0}^{N_{\rm sec}}\frac{1}{n}\Pr[N=n] <
	\Pr[N\leq n_0]+\frac{1}{n_0}\Pr[N> n_0]< \Pr[N\leq n_0]+\frac{1}{n_0},
\end{equation}
where $n_0$ is some positive integer less than expectation value $N_{\rm sec}/N_{\rm addr}$.
We then consider $n_0$ in the form $n_0:=kN_{\rm sec}/N_{\rm addr}$ with $k<1$.
According to Ref.~\cite{ProbEst}, the following upper bound on the binomial cumulative distribution function can be used:
\begin{equation}\label{eq:ub}
	\Pr[n\leq N_0]\leq \frac{N_{\rm sec}-k N_{\rm sec}/N_{\rm addr}}{N_{\rm addr}(N_{\rm sec}/N_{\rm addr}(1-k))^2}<\frac{N_{\rm addr}/N_{\rm sec}}{(1-k)^2}.
\end{equation}
By substituting Eq.~(\ref{eq:ub}) into Eq.~(\ref{eq:eps}) and choosing $k:=0.36$ providing an extremum value of RHS of Eq.~(\ref{eq:eps}), we obtain the following expression:
\begin{equation}
	\varepsilon < 5.22\times N_{\rm addr}/N_{\rm sec}<7\times10^{-29}<10^{-28}.
\end{equation}
One can see that the obtained value indeed is negligibly small.
Finally, we conclude that with probability larger than $1-\varepsilon$, that is extremely close to 1, the legitimate user would be able to provide the evidence of the brute-force attack.

\section{Diminishing attack motivation} \label{sec:modifications}

As we discussed above, the presence of a successful brute-force attack, considered in Sec.~\ref{sec:attack}, can be proven with the probability close to unity under reasonable assumptions about the employed cryptographic hash functions and users behavior.
However, there is a particular issue about evidencing the attack presence. 
The problem is that there is no straightforward way to figure out whose of two colliding public keys belongs to the adversary, and whose belongs to a legitimate user.
From the viewpoint of other users in the network both of the public keys seem to be equivalent.
We note that a time order in which these keys appear also does not resolve the problem.
One can think, that the first appeared public key is a forged one, but the second, shown as evidence, belongs to the legitimate user.
However, the adversary can wait until the legitimate user publishes his transaction and then disclose his public key as the evidence and claim that he suffered from the attack.

Here we consider a solution which aims not to distinguish between legitimate user and adversary, but to diminish the whole motivation of the brute-force attack.
This can be achieved by freezing funds transfer after an appearance of the evidence of the brute-force attack related to these funds.
The solution is based on introducing two modification of the current Bitcoin protocol.
The first modification consists of introducing a new transaction type which contains evidence of brute-force attack.
We call it an \emph{evidence transaction}.
The second modification is an introducing of timeout requirement between publishing a transaction and spending funds from its outputs.
This requirement is necessary to have enough time to publish the evidence transaction to the blockchain before the stolen funds will be spent (transferred to other probably legitimate users).
Let us consider the proposed modification in more detail.

\subsection{Evidence transaction}

The aim of the evidence transaction is to provide a possibility to present an alternative valid public key to another public key used to transfer funds from some \PtoPKH~transaction(s) output(s).
By alternative public key, we understand a public key which hashes to the same address which we call a \emph{disputed address}.
In contrast to standard cryptocurrency transactions, evidence transactions refer not to the outputs, but to the \PtoPKH~inputs of related transactions, which call as \emph{suspect transactions}.
The outputs of the suspect transaction have to be unspent, that should be provided by the second proposed modification.
The evidence transaction should contain alternative a public key corresponding to the published key in related inputs of suspect transactions, and also provide an auxiliary address, whose purpose will be discussed later.

If the evidence transaction got into blockchain, then the following operations should be performed.
\begin{enumerate}
	\item All the transactions that (i) contain \PtoPKH~inputs with public key colliding to public key presented in evidence transaction, and (ii) published in blocks appeared not earlier than the oldest block containing suspect transactions given in the evidence transaction, started to be considered as suspect transactions.
	\item All the outputs of the suspect transactions and all the outputs containing the disputed address are removed from \UTXO.
	\item Outputs of the transactions related to the inputs of the suspect transactions, which do not contain a disputed address, turn back into an unspent state (move back to \UTXO).
	\item An extra output is added to \UTXO, which contains a number of bitcoins given by the sum of funds on all the inputs of suspect transactions that have a colliding public key and fees of the suspect transactions.
	These funds should be able to be spent by publishing a transaction signed with the secret key corresponded to the public key in the suspect transaction (giving a disputed address), but can be only transferred to the auxiliary address given in the evidence transaction.
\end{enumerate}

The first three operations ensure that the funds taken from addresses which were not hacked continue their circulation in the system.
At the same time, the funds from disputed address will be frozen, and the only way how they can continue their circulation is their transfer by the author of suspect transactions to the address given by the author of the evidence transaction.
This scenario can be realized if the author of the suspect transaction is ``white hacker'', whose aim is only to demonstrate the vulnerability of the system, but not to steal funds.

An important issue related to evidence transaction is its fee to miners.
Since the legitimate author of the evidence transaction probably no longer has any coins after the suspect transaction is published, 
we propose to introduce an extra reward for miners to add evidence transactions to the blocks (the evidence transaction itself should bot contains any fee).
The reward could be calculated as a median value of total transaction fees in the last six blocks, and should be added to the number of coins emitted in the current block.

\subsection{Timeout requirement}

Since the outputs of a suspect transaction should not be spent by the time an evidence transaction is published, one need to introduce a timeout between publishing a transaction and spending its outputs in subsequent transactions.
This timeout is necessary for providing the opportunity for the legitimate user to react on unauthorized funds transfer and publish the evidence transaction.

In the Bitcoin network, it is recommended to wait until five blocks appear on top of the block containing a particular transaction in order to be (almost) sure that this transaction will not be removed from the blockchain.
From this perspective, the considered timeout can be selected at the same level of six blocks.
This choice, arguably, does not lead to a noticeable decrease in network performance.

Finally, we would like to emphasize that the considered approach of introducing an evidence transaction is applicable only in the case of the brute-force attack presented in Sec.~\ref{sec:attack}.
It is not valid in the case, where the adversary is able to find two valid secret keys giving a particular address at once.

\section{Rewarding the collision detection} \label{sec:reward} 

Another option for dealing with brute-force attacks, which does not require any changes in the present consensus, is introducing a reward for finding colliding public keys.
Obtaining the reward can be realized in automatic fashion by means of creating a corresponding smart contract.
It turns out that the Bitcoin script language allows creating such kind of rewarding.

Consider the following Bitcoin output script (\textsf{scriptPubKey}):

\medskip
\begin{tabular}{|l|}
	\hline
	\textsf{OP\_OVER} \textsf{OP\_OPVER} \textsf{OP\_EQUAL} \textsf{OP\_NOTIF} \textsf{OP\_OVER} \textsf{OP\_HASH160}\\
	\textsf{OP\_SWAP} \textsf{OP\_HASH160} \textsf{OP\_EQUALVERIFY} \textsf{OP\_CHECKSIG} \textsf{OP\_ELSE}\\
	 \textsf{OP\_RETURN} \textsf{OP\_ENDIF}\\
	\hline
\end{tabular}
\medskip

It can be interpreted as follows: ``The funds from the output can be redeemed by presenting a triple (\textsf{sig}$_1$, \textsf{pubkey}$_1$, \textsf{pubkey}$_2$), where \textsf{pubkey}$_1$ and \textsf{pubkey}$_2$ are two different public keys which give the same address, and \textsf{sig}$_1$ is a valid signature under the public key \textsf{pubkey}$_1$.
The fund from this output could be redeemed a legitimate user, whose funds were stolen, or by the adversary, who obtained some alternative public key.

The rewarding transaction can be published by anyone, who would like to participate in the project of detecting the brute-force attack.
Of course, the information about the presence of rewarding transaction has to be somehow spread among the community.

We also note that in presence of such rewards the adversary could include in his list of addresses for which he would try to find a valid secret key (\AddrList) all the addresses without funds but with public keys published in the blockchain.
Unfortunately, if there were some funds on address whose public keys were published, then the adversary could both steal the funds and obtain the reward.

Finally, we would like to make some remarks about why this kind of rewarding is valuable for the whole Bitcoin blockchain system. 
The first reason is providing some automatic compensation for a person who suffered from a successful brute-force attack.
The second and more important one is that in the case when someone redeems this reward, the whole network will obtain an important signal, that the employed cryptography (particularly) hash functions have an evident vulnerability or there is some participant (e.g. pool) which possess incredible computational resources.

\section{Conclusion} \label{sec:concl}

Here we summarize the main results of our work.
We have considered a brute-force attack on Bitcoin wallet which consists of finding secret keys for existing addresses.
We have demonstrated that if this attack succeeds then with a probability higher than $1-10^{-28}$ the legitimate user will be able to prove that it was the brute-force attack.
However, we also have shown that there is difficulty determining who is a legitimate user and who is an adversary.

We have considered two possible approaches to dealing with a possible brute-force attack.
The first approach involves modifications in present consensus and allows to freeze stolen funds transfer.
This approach allows diminishing attack motivation since even if the adversary will succeed in implementing the attack, with very high probability he will not be able to use the stolen funds.
This method may be used in developing a new cryptocurrency system.

The second approach does not involve any modification in the present Bitcoin consensus, but propose to create special reward transaction, which allows getting some coins as compensation to a person who suffered from brute-force attack.
More importantly, the fact of obtaining of this reward will serve as undeniable evidence the cryptography employed in the current blockchain network has some vulnerabilities.
This kind of reward transaction could be added by any person who wants to donate his coins to the project of detecting brute-force attacks.

\end{document}